\documentclass[11pt]{article}
\usepackage[letterpaper,top=1in,bottom=1in,left=1in,right=1in]{geometry}
\usepackage{times, amssymb}
\usepackage{authblk}

\usepackage{xcolor}
\usepackage[colorlinks=true,citecolor=blue,linkcolor=blue]{hyperref}
\let\emptyset\varnothing

\hypersetup{
    colorlinks,
    linkcolor={red!50!black},
    citecolor={blue!50!black},
    urlcolor={blue!80!black}
}

\usepackage[]{amsmath,amssymb,amsfonts,latexsym,amsthm,enumerate,enumitem,xcolor,cite,bbm}
\usepackage{algorithm2e}
\usepackage[nameinlink, noabbrev, capitalize]{cleveref}
\usepackage{comment}
\usepackage{nicefrac}
\crefname{prop}{Proposition}{Propositions}
\crefname{ineq}{inequality}{inequalities}
\crefname{proof}{Proof}{Proofs}
\creflabelformat{ineq}{#2(#1)#3}

\newcommand{\eqdef}{\stackrel{\textrm{def}}{=}}

\newtheorem{theorem}{Theorem}
\newtheorem{lemma}{Lemma}
\newtheorem{prop}{Proposition}
\newtheorem{proposition}{Proposition}
\newtheorem{claim}{Claim}
\newtheorem{fact}{Fact}

\newtheorem{question}{Question}

\newtheorem{corollary}{Corollary}
\newtheorem{definition}{Definition}

\crefname{THM}{Theorem}{Theorems}

\usepackage{subcaption}
\usepackage{graphicx}
\usepackage{tikz}
\usepackage{color, colortbl}
\definecolor{LightCyan}{rgb}{0.88,1,1}
\definecolor{Gray}{gray}{0.9}
\usetikzlibrary{positioning}
\usetikzlibrary{decorations.pathmorphing}
\usetikzlibrary{automata,positioning}
\usetikzlibrary{arrows}
\usetikzlibrary{arrows.meta}
\usetikzlibrary{calc}

\newcommand{\N}{\mathbb{N}}

\newcommand{\R}{\mathbb{R}}
\newcommand{\poly}{\operatorname{poly}}

\newcommand{\prob}{\mathbf{P}}

\newcommand{\zo}{\{0,1\}}

\newcommand{\Maj}{\mathsf{Maj}}

\newcommand{\cC}{\mathcal{C}}

\newcommand{\trn}{\tau}
\newcommand{\ntrn}{\#\Gamma}
\newcommand{\trans}{\Gamma}

\newcommand{\total}{\sigma}
\newcommand{\cutp}{\phi}

\newcommand{\openfunction}{Y}
\newcommand{\openinteger}{y}

\newcommand\enumballsat[2]{\textsc{Enum($#1$, $#2$)}}
\newcommand\enumz[1]{\textsc{Enumz($#1$)}}
\newcommand{\ts}{\textsc{TreeSearch}}
\newcommand\ballsat[2]{\textsc{($#1$, $#2$)-SAT}}

\usepackage{subfiles}
\newcommand{\dobib}{
    \bibliographystyle{alpha}
    \bibliography{refs} 
}

\begin{document}
\renewcommand{\dobib}{}

\title{Local Enumeration and Majority Lower Bounds}
\author[1]{Mohit Gurumukhani\thanks{E-mail: \texttt{mgurumuk@cs.cornell.edu}. Supported by NSF CAREER Award 2045576 and a Sloan Research Fellowship.}}
\author[2]{Ramamohan Paturi\thanks{E-mail: \texttt{rpaturi@ucsd.edu}. Partially supported by NSF grant 2212136.}}
\author[3]{Pavel Pudl{\'a}k\thanks{E-mail: \texttt{pudlak@math.cas.cz}. Partially supported by grant EXPRO 19-27871X of the Czech Grant Agency and the institute grant RVO: 67985840.}}
\author[4]{Michael Saks\thanks{E-mail: \texttt{saks@math.rutgers.edu}.}}
\author[3, 5]{Navid Talebanfard\thanks{E-mail: \texttt{n.talebanfard@sheffield.ac.uk}. This project has received funding from the European Union’s Horizon Europe research and innovation programme under the Marie Sk{\l}odowska-Curie grant agreement No 101106684 — EXCICO. Views and opinions expressed are however those of author(s) only and do not necessarily reflect those of the European Union or REA. Neither the European Union nor the granting authority can be held responsible for them.\\ \includegraphics[scale=0.1]{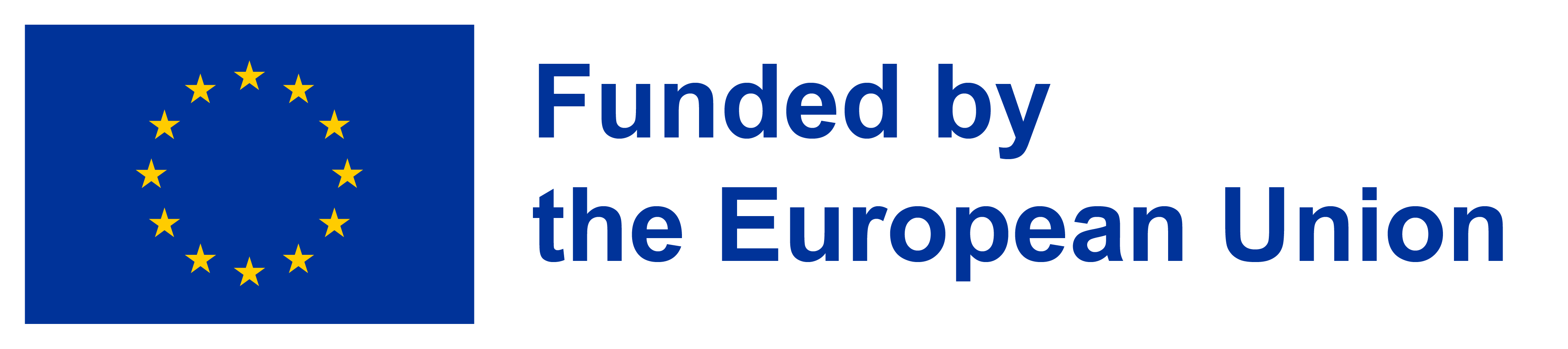}}}
\affil[1]{Cornell University}
\affil[2]{University of California, San Diego}
\affil[3]{Institute of Mathematics of the Czech Academy of Sciences}
\affil[4]{Rutgers University}
\affil[5]{University of Sheffield}


\maketitle

\begin{abstract}

Depth-3 circuit lower bounds and $k$-SAT algorithms are intimately related; the state-of-the-art $\Sigma^k_3$-circuit lower bound (Or-And-Or circuits with bottom fan-in at most $k$) and the $k$-SAT algorithm of Paturi, Pudl{\'a}k, Saks, and Zane (J. ACM'05) are based on the same combinatorial theorem regarding $k$-CNFs. In this paper we define a problem which reveals new interactions between the two, and suggests a concrete approach to significantly stronger circuit lower bounds and improved $k$-SAT algorithms. For a natural number $k$ and a parameter $t$, we consider the \textsc{Enum($k$, $t$)} problem defined as follows: given an $n$-variable $k$-CNF and an initial assignment $\alpha$, output all satisfying assignments at Hamming distance $t(n)$ of $\alpha$, assuming that there are no satisfying assignments of Hamming distance less than $t(n)$ of $\alpha$. We observe that an upper bound $b(n, k, t)$ on the complexity of \textsc{Enum($k$, $t$)} simultaneously implies depth-3 circuit lower bounds and $k$-SAT algorithms:
\begin{itemize}
    \item {\bf Depth-3 circuits:} Any $\Sigma^k_3$ circuit computing the Majority function has size at least $\binom{n}{\frac{n}{2}}/b(n, k, \frac{n}{2})$.

    \item {\bf $k$-SAT:} There exists an algorithm solving $k$-SAT in time $O(\sum_{t = 1}^{n/2}b(n, k, t))$.
\end{itemize}

A simple construction shows that $b(n, k, \frac{n}{2}) \ge 2^{(1 - O(\log(k)/k))n}$. Thus, matching upper bounds for $b(n, k, \frac{n}{2})$ would imply a $\Sigma^k_3$-circuit lower bound of $2^{\Omega(\log(k)n/k)}$ and a $k$-SAT upper bound of $2^{(1 - \Omega(\log(k)/k))n}$. The former yields an unrestricted depth-3 lower bound of $2^{\omega(\sqrt{n})}$ solving a long standing open problem, and the latter breaks the Super Strong Exponential Time Hypothesis.

In this paper, we propose a randomized algorithm for 
\textsc{Enum($k$, $t$)} and introduce new ideas to analyze it.
We demonstrate the power of our ideas by considering
the first non-trivial instance of the problem, 
i.e., \textsc{Enum($3$, $\frac{n}{2}$)}. 
We show that the expected running time of  our algorithm is 
$1.598^n$, substantially improving on the trivial bound of $3^{n/2} \simeq 1.732^n$. This already improves $\Sigma^3_3$ lower bounds for Majority function to $1.251^n$. The previous bound was $1.154^n$ which follows from the work of H{\r a}stad, Jukna, and Pudl{\'a}k (Comput. Complex.'95).

By restricting ourselves to monotone CNFs, \textsc{Enum($k$, $t$)} immediately becomes a hypergraph Tur{\'a}n problem. Therefore our techniques might be of independent interest in extremal combinatorics.

\end{abstract}



\section{Introduction} \label{sec:introduction}

Local search is a fundamental paradigm in solving the satisfiability problem: find an assignment close in Hamming distance to the initial assignment that satisfies the formula, if one exists. Papadimitriou \cite{Papadimitriou91} was the first to employ this idea in a randomized poly-time 2-SAT algorithm. Sch{\" o}ning \cite{DSchoning02} showed that a slight modification of this algorithm yields a running time of $(2 - 2/k)^n$ for $k$-SAT. Dantsin et al. \cite{DantsinGHKKPRS02} considered a deterministic version of local search and gave a deterministic $(2 - 2/(k+1))^n$ time algorithm. Brueggemann and Kern \cite{BrueggemannK04} and Kutzkov and Scheder \cite{KS} improved this deterministic local search procedure and obtained faster deterministic 3-SAT algorithms. Moser and Scheder \cite{MoserS11} eventually considered a variant of the local search problem and used it to give a deterministic $k$-SAT algorithm matching the running time of Sch{\" o}ning's.

Despite the success of local search, the fastest known $k$-SAT algorithm PPSZ and its improvements follow a different approach: pick a random variable $x$, if its value is not easily seen to be forced\footnote{For example if $x$ appears in a unit clause, or if such a clause can be derived in small width resolution.} then assign it randomly and continue (see \cite{PaturiPZ99,PaturiPSZ05,Hertli14,DBLP:conf/stoc/HansenKZZ19,Scheder21}). The analysis of this simple yet powerful algorithm consists of a combinatorial theorem relating the size and the structure of the set of satisfying assignments of a $k$-CNF. The strength of this combinatorial theorem is further manifested by the fact that the state-of-the-art depth-3 circuit lower bounds are built on it \cite{PaturiPZ99,PaturiPSZ05}.

This curious interaction between lower bounds and algorithms has become less of a surprise over the years. Williams \cite{Williams13} initiated a whole new line of inquiry by showing that improved satisfiability algorithms for a circuit class automatically imply lower bounds for the same class. Conversely, almost all known circuit lower bound techniques have been adopted in satisfiability algorithms (see e.g. \cite{ImpagliazzoMP12,ChenKKSZ15}). Within this context the role of local search is unclear.

\begin{question}
Can we derive lower bounds from local search algorithms?
\end{question}

This question is also motivated by a lack of progress
in improving depth-3 circuit lower bounds and 
related upper bounds on $k$-SAT algorithms.

\bigskip
\noindent{\bf{Depth-3 circuit lower bounds.}} A $\Sigma^k_3$ circuit is an 
Or-And-Or circuit where the bottom fan-in is bounded by $k$, i.e., a disjunction of $k$-CNFs. We use $\Sigma^k_3(f)$ to denote the minimum number of $k$-CNFs in a $\Sigma^k_3$ circuit computing a function $f$. The study of 
these circuits was advocated by Valiant \cite{Valiant77} who showed that a 
strong enough $\Sigma^k_3$ lower bound for every fixed $k$ implies 
a super-linear lower bound for series-parallel circuits. Moreover, \cite{GKW21circuit} showed that strong lower bounds even for small constant $k$ such as $k=16$ imply various circuit lower bounds including better general circuit lower bounds.
The technique of \cite{PaturiPZ99} gives a lower bound of $\Omega(2^{n/k})$ 
for the parity function and it is known to be 
tight. In fact this results in a $\Omega(n^{\frac{1}{4}} 2^{\sqrt{n}})$ lower bound
for computing parity by unrestricted 
depth-3 circuits which is tight up to a constant factor. 
A further improvement comes from  \cite{PaturiPSZ05} which 
gives a lower bound of $2^{cn/k}$ where $c > 1$ for the BCH code. 
At this point, this is the best known
lower bound for computing any explicit function by $\Sigma^k_3$ circuits.

\bigskip
Majority is a natural candidate 
for going beyond the $2^{\Omega(\sqrt{n})}$ depth-3 circuit lower bound.
The natural $\Sigma^k_3$ circuit for 
computing Majority has size $2^{O(n\log(k)/k)}$ (which implies an unrestricted depth-3 size upper bound of $2^{O(\sqrt{n\log n})}$). H\r{a}stad, Jukna and Pudl\'ak \cite{HastadJP95} introduced the intriguing notion of \emph{$k$-limits} to capture the depth-3 complexity of various functions  
and proved a lower bound of $2^{\Omega(\sqrt{n})}$ where the constant factor in the 
exponent is improved  over the constant one could obtain from Switching Lemma. 
Regarding $\Sigma^k_3$ circuits computing Majority, their result implies size lower bounds of  $1.414^n$ for $k = 2$, and $1.154^n$ for $k = 3$, and for $k \ge 4$ it yields nothing. The $\Sigma^2_3$ bound is known to be essentially tight \cite{PaturiSZ00}. More recently,
\cite{LecomteRT22} proved a tight lower bound of $2^{\Omega(n\log(k)/k)}$ for 
computing Majority by $\Sigma^k_3$ circuits where each 
And gate depends on at most $k$ variables.
Further, \cite{Amano23Majority} studied the effect of negations for $\Sigma^k_3$ circuits computing majority.
However, the question of proving tight lower bounds for computing Majority by depth-3 circuits 
(even for fan-in 3 circuits) remains open.


\bigskip
\noindent
{\bf $k$-SAT upper bounds:}
Lack of progress in improving the savings beyond $\Omega(\frac{1}{k})$
for $k$-SAT algorithms led researchers to consider
SSETH (Super Strong Exponential Time Hypothesis).
SSETH is the hypothesis that $k$-SAT cannot be solved with \emph{savings} asymptotically 
more than $1/k$, i.e., 
there is no $2^{(1 - \epsilon_k)n}$ time $k$-SAT 
algorithm with $\epsilon_k = \omega(1/k)$. 
However, SSETH is known to be 
false on average \cite{DBLP:conf/sat/VyasW19,LincolnY20}, 
that is, 
the satisfiability of almost all $k$-CNFs can be decided with 
much larger savings. 
It is thus not unreasonable to attempt to get such savings 
even in the worst-case. 
Yet we cannot hope to achieve such a savings for 
a large subclass of PPSZ-style algorithms \cite{DBLP:conf/coco/SchederT20}.

\bigskip
It appears that making progress towards larger $k$-SAT savings as well as depth-3 circuit lower bounds requires new ideas. We argue that local search has the potential to achieve this goal, and as an evidence for this claim, we apply local search ideas to give a new $\Sigma^3_3$ lower bound for Majority function.

\subsection{Local enumeration, $k$-SAT and $\Sigma^k_3$ lower bounds for Majority function}
The local search problem is formally defined as follows.

\bigskip
\noindent{\bf{\ballsat{k}{t}}.} Given an $n$-variable $k$-CNF $F$, a parameter $t$, and an assignment $\alpha$, decide if there is a satisfying assignment $\beta$ of $F$ such that $d(\alpha, \beta) \le t(n)$, where $d(\cdot)$ is the Hamming distance.

\bigskip
Dantsin et al. \cite{DantsinGHKKPRS02} gave a simple branching algorithm which solves \ballsat{k}{t} in time $\poly(n) \cdot k^t$. This already gives a non-trivial algorithm for 3-SAT: solve \ballsat{k}{\frac{n}{2}} 
starting with the all-0  and all-1 assignments. To get a non-trivial algorithm for larger $k$, they used \emph{covering codes}, i.e., a small and asymptotically optimal number $C(n, t)$ of Hamming balls of a given radius $t$ that cover the entire $n$-dimensional Boolean cube. Then an upper bound of $C(n, r)\cdot k^t$ follows immediately for $k$-SAT by solving \ballsat{k}{t} starting with 
the centers of each of the balls in the covering code. Setting $t = \frac{n}{k + 1}$ minimizes this quantity. 
Thus improved upper bounds 
for \ballsat{k}{t} immediately imply 
improved $k$-SAT upper bounds, 
and indeed this is what \cite{BrueggemannK04,KS} did  by
proving an upper bound of $c^t$ for some $c < 3$ for $\ballsat{3}{t}$.
However, this improvement in local search is not
sufficient to yield better upper  
bounds for $k$-SAT when we want to use the technique for large $t$. It is conceivable that improved bounds for large $t$ combined with covering codes would yield improved $k$-SAT algorithms. This leads to the following question:

\begin{question}
What is the complexity of \ballsat{k}{\epsilon n} where $0 < \epsilon \le \frac{1}{2}$?
\end{question}

It is also natural to consider the enumeration problem for \ballsat{k}{t}: 
enumerate all satisfying assignments within Hamming distance $t$ of an 
initial assignment. We note that even a weaker form of this problem 
already captures the circuit complexity of Majority function. For 
this purpose, we introduce the following class of parameterized problems.

\bigskip
\noindent{\bf{\enumballsat{k}{t}}.} Given an $n$-variable $k$-CNF $F$ 
and an initial assignment $\alpha$, 
output all satisfying assignments of $F$ at a Hamming distance $t$
from $\alpha$ assuming that there are no satisfying assignments
of $F$ at a Hamming distance of less than $t$ from $\alpha$.

\bigskip
We observe that upper bounds on \enumballsat{k}{t} imply depth-3 circuit lower bounds and $k$-SAT algorithms.

\begin{proposition}
Assume that \enumballsat{k}{t} can be solved in randomized expected time $b(n, k, t)$. Then 
\begin{enumerate}
	\item any $\Sigma^k_3$ circuit requires at least $\binom{n}{n/2}/b(n, k, \frac{n}{2})$ size for computing Majority function.
	\item $k$-SAT can be solved in time $O(\sum_{t = 1}^{n/2}b(n, k, t))$.
\end{enumerate}
\end{proposition}

\begin{proof}
1) Consider a $\Sigma^k_3$ circuit $C$ that computes Majority function.
We will write $C= \bigvee_{i = 1}^m F_i$, where each $F_i$ is a $k$-CNF. 
None of the $F_i$s has a satisfying assignment with Hamming 
weight less than $n/2$. 
By assumption we can enumerate all satisfying assignments of Hamming weight exactly $n/2$ in expected time $b(n, k, \frac{n}{2})$. 
This in particular implies that the total number of such 
satisfying assignments for each $F_i$ is at most $b(n, k, \frac{n}{2})$. 
Since $F_i$s should together cover all 
assignments of Hamming weight exactly 
$n/2$ and since there are $\binom{n}{n/2}$ such assignments, 
the claim follows.

2) We can trivially check if there is a satisfying assignment of Hamming weight at most $n/2$  in $1 + \sum_{t = 1}^{n/2}b(n, k, t)$ steps. In the same number of steps we can check if there is a satisfying assignment of Hamming weight at least $n/2$.
\end{proof}

Observe that $b(n, k, \frac{n}{2})$ cannot be too small: 
define the $k$-CNF $\Maj_{n, k}$ by partitioning the $n$ variables into 
sets of size $2(k - 1)$ and 
by including all positive clauses of size $k$
from each of the parts. It is easy to see that every satisfying assignment 
of this formula must set at least $k - 1$ variables in each 
part to 1 and the total number satisfying assignments with 
Hamming weight $n/2$ is $2^{(1 - O(\log(k)/k))n}$, 
thus $b(n, k, \frac{n}{2}) \ge 2^{(1 - O(\log(k)/k))n}$. It 
follows that a matching upper bound 
for \enumballsat{k}{\frac{n}{2}} refutes SSETH and 
gives $\Sigma^k_3$ lower bound of $2^{\Omega(\frac{\log k}{k}n)}$ 
for Majority which in turn implies a $2^{\omega(\sqrt{n})}$ unrestricted depth-3 circuit lower bound, breaking a decades long barrier.

\subsection{Our contributions}

In this paper, we study \enumballsat{3}{\epsilon n} 
and obtain new algorithms and lower bounds. Note that this is the first non-trivial instance of \enumballsat{k}{t}, since \enumballsat{2}{t} can be solved in $2^t$ steps using a simple extension of the local search algorithm, and it is easy to see that this is tight for $t \le n/2$ by considering the 2-CNF consisting of $t$ disjoint monotone clauses.

\begin{theorem}[Main result]
\label{thm:main-result}
\enumballsat{3}{t} can be solved in expected time

\begin{enumerate}
	\item $3^t$, for $t \le \frac{n}{3}$,
	\item $1.164^n \times 1.9023^t$, for $\frac{n}{3} < t \le \frac{3n}{7}$,
	\item $1.1962^n \times 1.7851^t$, for $\frac{3n}{7} < t \le \frac{n}{2}$.
\end{enumerate}

In particular, \enumballsat{3}{\frac{n}{2}} can be solved in expected time $1.598^n$. 
\end{theorem}

Consequently, we get
\begin{corollary}
$\Sigma^3_3(\Maj) \ge 1.251^{n - o(n)}$.
\end{corollary}

Our lower bound is the best known bound
to compute Majority function by $\Sigma^3_3$ circuits.
Note that $\Maj_{n, 3}$ has $6^{n/4} \simeq 1.565^n$ satisfying assignments of Hamming weight $n/2$. Our bound is not too far from the optimal bound 
and it is a substantial improvement over $3^{n/2} \simeq 1.732^n$.

In the following, we explain how our approach to enumeration differs 
from the well-known approaches.
The $k^t$ algorithm used by \cite{DantsinGHKKPRS02} which 
solves \ballsat{k}{t} is a simple branching procedure. Without 
loss of generality assume that the initial assignment is all-0. If 
there is no monotone clause in the formula, 
then the all-0 assignment satisfies the formula. 
Otherwise select a monotone clause $C = x_1 \vee \ldots \vee x_k$. 
Then for each $x_i$, recursively solve \ballsat{k}{(t-1)} 
for the formula restricted by $x_i = 1$. 
An obvious weakness of this algorithm is that 
if the depth of the recursion tree is more than $n/k$, 
then some assignments will be considered more than once in 
the tree thus leading to redundant computation.

\bigskip
Our starting point is to search
the recursion tree 
(which we call a transversal tree following the hypergraph nomenclature) 
so each satisfying assignment (which we call a transversal in the following) within the ball is visited exactly once.
We observe that such a non-redundant search can be conducted 
with any clause ordering and any fixed ordering of variables 
within the clause. During the search of the transversal tree, 
it is easy to decide whether 
a subtree contains any new satisfying assignments by considering 
the labels of the child edges of the nodes along the path.
If there are no new satisfying assignments in a subtree,  we will prune 
it.
It turns out that this approach is isomorphic to the 
seminal method of 
Monien-Speckenmeyer \cite{MonienS85} where for each $i$ we 
recursively solve the problem under the 
restriction $x_1 = \ldots = x_{i - 1} = 0, x_i = 1$. 
However, it is not clear
how to improve 
upon the bound obtained by \cite{MonienS85}.
We show that by choosing 
the clause ordering  carefully and randomly ordering 
clause variables, a better bound can be obtained. 
In other words, we will consider 
\emph{randomized} Monien-Speckenmeyer trees with careful clause ordering.
The crux of our contribution is a  new 
analysis of randomized transversal trees.

\bigskip
\noindent{\bf{Connection to hypergraph Tur{\'a}n problems.}} 
Recall that a transversal in a hypergraph is a set of 
vertices that intersects every hyperedge. 
The recent work \cite{FranklGT22} gives a connection 
between depth-3 circuits and 
transversals\footnote{\cite{FranklGT22} results 
are stated in terms of cliques which are dual to transversals.}. 
Here we find 
another connection. Given $n$, $t$, and $k$, 
let $R^+(n, t, k)$ be the maximum number of 
transversals of size $t$ in an $n$-vertex $k$-graph 
with no transversal of size $t - 1$. 
Since a $k$-graph can be viewed as a monotone $k$-CNF, our algorithm 
enumerates minimum size transversals and thus gives new upper 
bounds for $R^+(n, t, k)$. 
Mantel's theorem, which is a special case of 
Tur{\'a}n's seminal theorem, gives the 
exact value $R^+(n, n - 2, 2) = n^2/4$, 
and the Tur{\'a}n problem for 
3-graphs can be phrased as 
showing $R^+(n, n - 3, 3) = (5/9 + o(1))n^3$ 
(see \cite{MR2866732} for a thorough survey of this 
and related problems). Our technique allows us to derive 
new bounds for $R^+(n, t, 3)$, 
where $t = \Theta(n)$. 
Although we currently do not get any useful results 
for  $t = n - o(n)$, we hope that our techniques can be extended to make progress for
this regime of parameters.

\bigskip
\noindent{\bf{Enumeration algorithms for CNFs with bounded negations.}} 
As an additional application of our enumeration techniques, we get the following enumeration algorithm:
\begin{theorem}
\label{thm:enum3negate}
Let $F$ be a CNF of arbitrary width where either each clause contains at most $3$ negative literals or each clause contains at most $3$ positive literals. Then, we can enumerate all minimal satisfiable solutions of $F$ in time $O(1.8204^n)$.
\end{theorem}

\section{Preliminaries} \label{sec:preliminaries}

In this section, we introduce concepts and 
notation that we use for the rest of the paper.
Let $F = (X, \cC)$ be a $k$-CNF with variable set $X$ and clause set $\cC$. We view the satisfying assignments of $F$ as 
subsets of variables which are set to 1.

\begin{definition}[Transversals]
A set $S\subseteq X$ is a transversal for $F$ if the assignment that sets
the variables in $S$ to $1$ and the variables in 
$X\setminus S$ to $0$ is a satisfying assignment of $F$.
The size of a transversal $S$ is defined as $|S|$.
\end{definition}

We say $S$ is a minimal transversal for $F$ if no subset of $S$ is a transversal. We call the corresponding satisfying asssignment a minimal satisfying assignment.

Our use of transversals for discussing satisfying assignments is
motivated by the notion of transversals in hypergraphs.
We view a monotone 
$k$-CNF 
(where all literals in every clause are positive) 
as a $k$-graph where every hyperedge has size at most $k$. 
This leads to a 1-1 correspondence
between the transversals of a monotone
$k$-CNF 
and those  of the corresponding hypergraph.

In this paper, we are primarily interested in 
minimum-size transversals.

\begin{definition}[Transversal number]
For a satisfiable $k$-CNF $F$, we define \emph{transversal number} $\trn(F)$ to be the cardinality of the minimum-size transversal of $F$.  
We use $\trans(F)$ to denote the set of all minimum-size
transversals of $F$
and $\ntrn(F)$ to denote the cardinality of 
$\trans(F)$.
\end{definition}

Let $t\in [n]$ and $\alpha\in \zo^n$ be such 
that every satisfying assignment of $F$ is at a distance
of at least $t$ from $\alpha$. 
We reduce the $\enumballsat{k}{t}$  problem for $F$
from the initial assignment $\alpha$
to the $\enumz{k}$  problem:
enumerate all 
minimum-size transversals of a $k$-CNF\footnote{This problem has been previously considered by, e.g. \cite{FominGLS19} with a different name {\sc Min-Ones $k$-SAT} and a non-trivial algorithm is given independent of $\trn(F)$. Here we focus on running times with fine dependence on $\trn(F)$.}.
Indeed, let $G$ be the $k$-CNF formula (over 
the variable set $X$) where its
clause set is obtained from that of
$F$ by the following replacement of literals:  
For each variable $v\in X$, if $\alpha$ sets $v$ to $1$, 
then we swap occurrences of the positive and the negative 
literals corresponding to $v$ in $\cC$. 
Otherwise, if $\alpha$ sets $v$ to $0$,  we leave the corresponding literals
as they are.
$G$ is also a $k$-CNF and 
for all $y\in \zo^n$, $G(y) = F(y\oplus \alpha)$. Clearly, 
there exists a transversal within distance $t$ 
from $0^n$ in $G$ if and only if there exists a transversal 
within distance $t$ from $\alpha$ in $F$. 

We now prove the following useful proposition that
allows us to assume all clauses of $F$ have width exactly $k$.

\begin{proposition}
\label{prop:monotone-k-cnf-uniform}
For every $k$-CNF $F = (X, \cC)$ with $\trn(F) \le n-k$,
there exists a $k$-CNF $G$ where every clause has width exactly $k$
and $\trn(G) = \trn(F)$
and the set of transversals of $G$ includes the
set of transversals of $F$.
Furthermore, if $F$ is monotone, $G$ is monotone.
\end{proposition}

\begin{proof}
Assume that $F$ has a clause $C$ of width $1\leq k'<k$.
Let $F'$ be the formula obtained from $F$ by removing the clause $C$
and adding a clause  $C' = C\cup C'' $ for
each $S\subseteq X\setminus C, |S|= k-k'$, where $C''$ is a monotone clause over variables in $S$.
The proposition follows since every transversal of
$F$ is a transversal of $F'$ and any transversal
of $F'$ which is not a transversal of $F$ must have size at least $n-k+1$.
\end{proof}

\section{Transversal Trees  and Tree Search} \label{sec:transversal-tree}

In this section, we present an algorithm called $\ts$ 
for solving $\enumz{k}$, i.e., 
to enumerate all minimum-size transversals 
of a satisfiable $k$-CNF $F$. Let  $t=\trn(F)$ be the transversal number of $F$.
Our algorithm considers a tree,
called \emph{transversal tree}, of depth $t$
where each minimum-size transversal corresponds to at least
one leaf node at depth $t$.
Our algorithm $\ts$ traverses the tree to enumerate 
the leaves at depth $t$ corresponding to 
minimum-size transversals. 
However, 
a leaf at depth $t$ 
need not correspond to a minimum-size transversal and 
furthermore two distinct leaves at depth $t$ may 
correspond to the same
minimum-size transversal.
For these reasons, 
enumerating all leaves can take significantly more
time than the total number $\ntrn(F)$
of minimum-size transversals.
We
deal with this issue by pruning subtrees so that 
$\ts$ would only encounter 
minimum-size transversals 
that are not encountered elsewhere.
While our pruning approach is isomorphic to that of Monien-Speckenmeyer \cite{MonienS85}, 
our analysis and bound crucially depend on 
our choice of clause ordering 
and random ordering of the child nodes of the transversal tree.
In the following, we define the required concepts and present 
our constructions.

\begin{definition}[$(k,X)$-trees]
For $k\geq 1$ and  a set $X$ of variables, a $(k,X)$-tree
is a directed $k$-ary tree $T$ 
with a node and edge labeling $Q$ and a tree-edge ordering $\pi$
which satisfies the following properties.
\begin{enumerate}
\item Each edge  is directed from parent to child.
Each non-leaf node has at most $k$ children. Children of a node
are ordered from left to right according to $\pi$.
\item 
Each (tree) edge $e$ is labeled with a variable $q_e\in X$.
Each node $v$ is labeled with a set $Q_v \subseteq X\cup\{\bot\}$.

\item 
For each node, labels of child edges are distinct.
If $e= (u,v)$ is an edge, then 
$Q_v = Q_u\cup \{q_e\}$ or
$Q_v = Q_u\cup \{q_e, \bot\}$.
\item
Labels of  edges along any path are distinct.
\item All leaves $v$ where $\bot\notin Q_v$ 
are at the same level.
\end{enumerate}
\end{definition}
Let $T$ be a $(k,X)$-tree with root $r$ and labeling $Q$.
For node $v$ of $T$, 
let $T_v$ denote the subtree $T_v$ of $T$ rooted at $v$.
If $u$ and $v$ are nodes of $T$ 
such that $u$ is an ancestor of $v$, $P_{uv}$ denotes the unique 
path from $u$ to $v$ in $T$.


\begin{definition}[Shoot of a tree path]
If $u$ and $v$ are nodes of $T$ such that $u$ is an ancestor of $v$, the subgraph consisting of all the child edges of the nodes 
along the unique path from $u$ to $v$ is called the \emph{shoot} $S_{uv} = S^T_{uv}$ from $u$ to $v$. In particular $P_{uv} \subseteq S_{uv}$.
\end{definition}

For a path $P_{uv}$, the labels of the  edges along the path are called \emph{path variables} of $P_{uv}$.
For a shoot $S_{uv}$, labels of the shoot edges are called shoot variables.

\begin{definition}[Transversal tree]
Let $F=(X,\cC)$ be a $k$-CNF on the variable set $X$.
A $(k,X)$-tree $T$ rooted at $r$ with labeling $Q$ and tree-edge ordering $\pi$
is a \emph{transversal} tree for $F$ if
\begin{enumerate}
\item $Q_r = \emptyset$ if $F$ has no empty clause 
and otherwise $Q_r = \{\bot\}$, and 
\item for every node $v$, each 
minimum-size transversal of $F$ which is an 
extension of $Q_v$ will appear as 
the label of a leaf in the subtree rooted at $v$.
\end{enumerate}
\end{definition}

It is easy to see that  a transversal tree $T$ for a satisfiable
$k$-CNF $F$ has depth $\trn(F)$ and 
every leaf $v$ of $T$ such that $\bot \notin Q_v$ is at depth $\trn(F)$.
We also note that any subtree of a transversal tree is also a transversal tree.

\begin{definition}[Valid and invalid leaves]
Let $T$ be a transversal tree for a satisfiable $k$-CNF $F$.
We say that a  leaf $v$ of $T$
is \emph{valid} if $Q_v$ is a minimum-size transversal of $F$.
Otherwise it is \emph{invalid}.
\end{definition}

For a transversal tree $T$ of a satisfiable $k$-CNF $F$, 
let $\trans(T)$ denote the 
collection  of minimum-size transversals associated 
with the valid leaves of $T$. The definition of transversal tree implies the following basic fact.

\begin{fact}
$\trans(F) = \trans(T)$.
\end{fact}

\subsection{Construction of Transversal Trees}

In this section, we show how to construct transversal trees for 
a satisfiable $k$-CNF 
$F=(X,\cC)$. 
The construction produces a labeling $Q$ on nodes and edges.
We will not specify a tree-edge ordering in the construction.
However, we will later select
a  left-right ordering where child nodes 
are ordered randomly and independently for each node.
The construction depends on the ordering of clauses.
Let $\Pi$ denote an ordering of the clauses in $F$.

We start with the tree $T$ with just one node $r$
(the root node) with the label  $Q_r =\emptyset$.
Assume that we are about to expand a non-leaf node $v$.
By construction, $v$ is at depth less than $\trn(F)$ and $\bot \not\in Q_v$.
Since $v$ is at depth less than $\trn(F)$, $Q_v$ is not a transversal.
Select the first monotone clause 
$C_v = \{a_1, \ldots, a_{k'}\}$ according to the clause order $\Pi$, where $k'\le k$.
Such a monotone clause must 
exist since every clause is non-empty and since
otherwise an all-0 assignment will satisfy the formula
contradicting the fact that $Q_v$ is not a 
transversal. 
Also, it must be the case that 
$C_v\cap Q_v = \emptyset$ as none 
of the variables from $Q_v$ can appear in $C_v$.
For each $a\in C_v$,
\begin{enumerate}
\item Create a child node $v_a$ for $v$ and label the edge $(v, v_a)$ by $a$. 
\item Simplify the clauses 
by setting the variables
along the path $P_{rv_a}$ to 1. 
If there is an empty clause, set $Q_{v_a} = Q_v\cup\{a,\bot\}$ 
and $v_a$ 
will not be expanded and thus will become a leaf node.
Otherwise,
label $v_a$ by $Q_{v_a} = Q_v\cup \{a\}$. 
\item If the level
of the node is $\trn(F)$, make it a leaf node.
\item Order the child nodes of $v$ left-right according to a 
tree-edge ordering.
\end{enumerate}

\begin{proposition}
\label{prop:general-tree-existence}
The tree $T$ with the labeling as described above is a transversal tree for $F$.
\end{proposition}

\begin{proof}
Indeed each non-leaf node has at most $k$ children.
All leaves $v$ with $\bot\notin Q_v$ must be at the same level $\trn(F)$
since any node $v$ at a level smaller than $\trn(F)$ 
and does not contain $\bot$ in $Q_v$ can be expanded
and since the construction stops at level $\trn(F)$.
It is easy to verify that the labeling $Q$ has the requisite properties.
For every $v$ with $\bot \notin Q_v$
and every extension $Y$ of $Q_v$ to a minimum-size transversal, there exists a
leaf with label $Y$ in the subtree $T_v$ since there is a child edge $(v,v')$ of $v$ with label $a$ for some $a \in Y \cap C_v \ne \emptyset$ where $C_v$ is the clause used to expand $v$. Inductively we can construct a path from $v'$ to a leaf which ultimately has $Y$ as the label.
\end{proof}

\subsection{$\ts$: An Algorithm for Enumerating the Valid Leaves of $T$}

Our goal is to search the transversal tree to enumerate 
all minimum-size transversals. 
However, visiting 
all leaf nodes
may take at least $k^{\trn(F)}$ time. 
To improve the efficiency of the search,
we prune the tree during our search while
guaranteeing the enumeration of  
each minimum-size transversal exactly once.


Let $F$ be satisfiable $k$-CNF.
Let $\Pi$ be a clause ordering for $F$. Let $T$ be transversal
tree for $F$ constructed using $\Pi$ and some tree-edge ordering $\pi$. 
We note that for any tree-edge ordering $\pi$,
the edges of any 
shoot $S_{uv}$ (where $u$ is any ancestor of $v$)
are situated in one of three ways with respect to
the tree path from $u$ to $v$:
1) to the left of the tree path, 
2) to the right of the tree path, or
3) along the tree path.
Our key insight is that for any node $v$
we can determine 
whether the  subtree $T_v$ potentially contains 
any \emph{new }
minimum-size transversals by considering the labels of the edges
in the shoot $S_{rv}$.

Our $\ts$ starts with the root node $r$ of $T$.
Assume that we are currently visiting the node $v$.
Let $T_v$ be a subtree of $T$ rooted at $v$.
If $v$ is not a leaf, let $C_v$ be the monotone clause used for expanding
the node $v$ (based on the clause order $\Pi$) and   $a_1, \ldots, a_{k'}$ be the ordering of its variables
according to $\pi$ for some $k' \le k$. For $1\leq i \leq k'$, 
let $v_{a_i}$ be the $i$-th child node of $v$. The edge $e_i=(v,v_{a_i})$ is labeled with $a_i$.
The search procedure $\ts$ starting at node $v$ works as follows:

\begin{itemize}
\item  If $v$ is a leaf,
output $Q_v$ if it is a transversal.
In any case, return to the parent.
\item Otherwise, process the children of $v$ in order. 
Let $T_{i} = T_{v_{a_i}}$ be the transversal tree 
rooted at the child $v_{a_i}$ for $a_i \in C_v$. 
Prune the subtree $T_i$ if and only if 
the shoot $S_{rv}$ contains an edge $e'=(u',v')$ 
where $u'$ is ancestor of $v$ such that $Q_{e_i} = Q_{e'}$, and
the edge $e'$ appears to the left of the path $P_{rv}$.
Search the tree $T_i$ if it is not pruned.

\end{itemize}

\begin{fact}
For any clause ordering $\Pi$
and tree-edge ordering $\pi$, 
$\ts$ outputs all minimum-size transversals of $F$ exactly once.
\end{fact}

\noindent
\subsection{Canonical Clause Ordering and Random Tree Edge Ordering}

The time complexity of $\ts$ is bounded by the number of leaf nodes it visits.
While we know that $\ts$ outputs 
all minimum-size transversals without redundancy, 
it is much less clear how to analyze its complexity.
We need two ideas to analyze $\ts$ to get a good bound.
The first idea is a \emph{canonical} clause ordering  $\Pi$
in which a sequence of  maximally disjoint
monotone clauses will precede all other clauses.
The second idea is a random $\pi$, that is, 
a tree-edge ordering that orders 
the children of every node in the transversal tree
uniformly 
and independently at random.


\section{Analysis of $\ts$ for Monotone $k$-CNFs} \label{sec:monotone}

Let $F=(X,\cC)$ be a monotone $k$-CNF where every clause has exactly $k$ literals.
Let $t=\trn(F)$ be the transversal number of $F$. We assume 
that $t\leq \frac{3n}{5}$.
Let $T$ be a transversal tree for $F$ where
$T$ is constructed using a canonical clause ordering $\Pi$.
The child edges of each of its nodes are randomly ordered from left-right independent of other nodes. Let $\pi$ denote this random tree-edge ordering. Let $r$ be its root and $Q$ its labeling.

In this section, we analyze 
$\ts$  
and prove \cref{thm:main-result} for 
the monotone case. We start
with a few ideas required to keep track of the effect of the random ordering on pruning.
We then build upon them 
\cref{sec:non-monotone} to prove \cref{thm:main-result} for general $k$-CNFs.

\subsection{Random Tree Edge Ordering and Pruning}

\begin{definition}[Cut event]
We say that an edge $e=(u,v)$ is \emph{cut} if  
$u$ has an ancestor $u'$ with a child edge $e'= (u',v')$
where $Q_{e'}=Q_e$ and 
$e'$ appears to the left of the path $P_{ru}$ according to $\pi$.
\end{definition}

We use $\cutp(e)$ to denote the event that the edge $e$ is \emph{not} cut.
We also use $\cutp(P_{uv})$ to denote the event no edge along $P_{uv}$ is cut.
We use $\cutp(u)=\cutp(P_{ru})$ to denote the event that none of the edges along the path from the root to the node $u$ are cut. 

\begin{definition}[Survival probability of paths and nodes]
The survival probability $\sigma(P_{uv})$
of a path $P_{uv}$ is $\prob(\cutp(P_{uv}))$.
The \emph{survival probability} $\sigma(u)$ of a node $u$ is $\prob(\cutp(u))$.
\end{definition}

\begin{definition}[Survival value of a transversal tree]
For a subtree $T_u$ rooted at $u$,
we define
$\total(T_u) = \sum_{\text{$v$ is leaf of $T_u$}} \sigma(v)$
as the survival value of $T_u$. We write $\total(T) = \total(T_r)$ where 
$r$ is the root node.
\end{definition}

Our main tool for upper bounding the expected running time of $\ts$ is the following basic fact.

\begin{fact}
\label{fact:ntrans-tree}
The expected number of leaves visited by $\ts$ is exactly $\total(T)$. In particular $\ntrn(F) \leq  \total(T)$.
\end{fact}

\begin{proof}
This follows by definition, since $\ts$ only visits surviving leaves of $T$ under a random tree-edge ordering. Furthermore, let $Y$ be  a minimum-size transversal of $F$.
We argue that $\sum_{v:Q_v=Y }\prob(\pi(P_{rv})) = 1$ which implies 
$\ntrn(F) \leq  \total(T)$.
\end{proof}

Our goal is to upper bound $\total(T)$ by bounding 
the survival probabilities of the leaves at depth $t$. 
A path survives if and only none of its edges are cut.
For an edge to be cut, it is necessary that some ancestor of the edge 
has a child edge with the same label as that of the edge. 
We will keep track of repeated edge labels via markings to bound 
the cut probabilities from below and thereby bounding the survival probabilities
from above.

\begin{definition}[Marking set of an  edge]
The \emph{marking set} $M(e)$ of an edge $e=(u,v)$ in $T$ is the set of
nodes $w\neq u$ in the path $P_{ru}$ which have a child edge $e'$ such that $Q_e=Q_{e'}$.
We say that the nodes in $M(e)$ mark the edge $e$.
We also say that the nodes in $M(e)$ mark the label of $e$. 
\end{definition}

\begin{definition}[Marked edges]
An edge $e=(v,u)$ in $T$ is \emph{marked}
if $M(e)\neq \emptyset$. 
\end{definition}

Marked edges are precisely those that have a non-zero probability of being cut. In fact, we can calculate the survival probability exactly.

\begin{fact}
For an edge $e$ in $T$,  $\sigma(e) = 2^{-|M(e)|}$.
\end{fact}

\begin{definition}
Let $u$ be a node in $T$. For each node $v$ 
along the path $P_{ru}$, let $N_u(v) = \{e\in P_{ru} \mid v\in M(e)\}$.
$N_u(v)$ is the set of edges along the path $P_{ru}$ marked by $v$.
\end{definition}

\begin{fact}
\label{fact:path-prob}
For any node $u$ in $T$,
the survival probability $\sigma(P_{ru})$ of the path $P_{ru}$
is given by 
\begin{align*}
\sigma(P_{ru}) = \Pi_{v: \text{ a node along }  P_{ru}} \frac{1}{|N_u(v)|+1}
\end{align*}
\end{fact}

\begin{proof}
$P_{ru}$ survives if and only if for every node $v \in P_{ru}$ and every 
edge $e \in P_{ru}$ that $v$ marks, 
the child edge of $v$ with the same label as that of $e$ appears to the right of the path. This happens with probability $\frac{1}{|N_u(v)| + 1}$ and since these events are independent, the claim follows.
\end{proof}

\subsection{An Analysis of $\ts$ for Monotone $3$-CNF}

Although \cref{fact:path-prob} gives us a fairly complete picture of 
the survival probabilities of individual paths in $T$ in terms of the edge
markings of the path,  we need a few ideas to find nontrivial upper bounds on $\total(T)$.
The first idea is the concept of a weight which captures the number of 
marked edges in a path or a shoot. 

\begin{definition}[Weight]
Let $P_{uv}$ be a path in $T$. 
The \emph{weight} of $P_{uv}$ is defined as 
$W(P_{uv}) \eqdef |\{e \in P: M(e) \ne \emptyset \}|$, 
i.e., the number of marked edges along $P_{uv}$. 
The \emph{weight} of a shoot $S_{uv}$ denoted by $W(S_{uv})$ is the number of marked edges 
in the shoot $S_{uv}$.

\end{definition}

The following fact provides a lower bound on the weight of each root to leaf
shoot  in $T$.

\begin{fact}
\label{fact:shoot-extended-weight-lower-bound}
Every root to leaf shoot $S_{ru}$ in $T$ has a weight of
at least $3t - n$.
\end{fact}

\begin{proof}
Since the depth of $T$ is $t$,
a root to leaf shoot has $3t$ edges and there are only $n$ distinct 
edge labels, at least $3t - n$ labels appear at least twice.
\end{proof}

\begin{definition}[Weight of a tree]
We say that a tree has weight $w$ if every root to leaf shoot of the tree has weight at least $w$.
\end{definition}

\begin{definition}[$M(w, d)$]
For non-negative integers 
$w$ and $d$, 
let $M(w, d)$ be the maximum survival value
over all ternary depth-$d$ transversal trees with weight $w$.
\end{definition}

We can now upper bound $\total(T)$ in terms of $M(w,d)$ 
exploiting the canonical clause ordering.
Our canonical clause ordering $\Pi$ starts with a 
maximal collection of disjoint clauses $C_1, C_2, \cdots, C_m$
so that all clauses in the formula intersect with at least one of the clauses from these disjoint clauses. 
We observe that $m\geq \frac{t}{3}$ for monotone $F$
since otherwise by setting all the variables in the monotone 
clauses $C_i$, we satisfy $F$ contradicting 
that the transversal number of $F$ is $t$.

\begin{lemma}
\label{pess-ntran}
\[
\total(T) \le
\begin{cases}
3^t & t \le \frac{n}{3}\\
3^{\frac{t}{3}}\times M(3t - n, \frac{2}{3}t) & \textrm{ otherwise}
\end{cases}
\]
\end{lemma}

\begin{proof}
For $t\le \frac{n}{3}$, we use the trivial upper 
bound of 1 on the survival probability of paths
to conclude that $\total(T) \le 3^t$ as desired.

For $t \ge \frac{n}{3}$, we use the fact that the canonical clause
ordering starts with a maximal collection of disjoint clauses $C_1, C_2, \cdots, C_m$
where  $m\geq \frac{t}{3}$.
We observe that for $1\leq i\leq \frac{t}{3}\leq m$ 
each node at level $i$ of $T$ is expanded by the same
clause $C_i$.
Moreover, none of  the child edges  of nodes at level $1\leq i\leq \le \frac{t}{3}\leq m$ are marked
as the corresponding clauses are disjoint.
The result follows since
for every node $u$ at level $\frac{t}{3} + 1$, the subtree $T_u$ has depth $\frac{2t}{3}$ and
the weight of
every root to leaf shoot in $T_u$
is at least $3t - n$ minus the number of marked edges from root to $u$ = $3t - n - 0 = 3t - n$ 
(\cref{fact:shoot-extended-weight-lower-bound}).
\end{proof}

\subsubsection{Upper Bounds on $M(w,d)$}
Upper bounding $M(w,d)$ for $T$ based on random tree-edge ordering $\pi$
is challenging. 
Instead we introduce a different
random process $\pi'$ for $T$: Each edge $e$ survives with probability $p_e$ independently
where $p_e = \lambda$ if $e$ is marked and $1$ otherwise,
where we define 
$\lambda \eqdef \frac{1}{\sqrt{3}}$.
The concept of survival probability under $\pi'$
can be extended to paths and nodes of $T$.
For example, the $\sigma'(P) = \prod_{e\in P} p_e$ is the survival probability
of the path $P$ under $\pi'$. 
Similarly, we define the survival value $\total'(T)$ of the transversal tree $T$ according to $\pi'$ as $\sum_{v\text{ is a leaf}} \sigma'(P_{rv})$. 
We define $M'(w,d)$ as the maximum survival value $\total'(T')$ of transversal trees $T'$
of depth $d$ where every root to leaf shoot has weight at least $w$.
The following lemma shows that $\total(T)\leq \total'(T)$.

\begin{lemma}
\label{path-pessim}
For a root to leaf path $P_{ru}$, 
$\sigma(P) \le \lambda^{W(P_{ru})} = \sigma'(P)$ which in turn implies $\total(T)\leq \total'(T)$
and $M(w,d)\leq M'(w,d)$. 

\end{lemma}

\begin{proof}
Given an edge $e \in P_{ru}$, define the contribution $y_e$ of $e$ to the
	survival probability (according to $\pi$) of $P_{ru}$ as 
\begin{align*}
y_e \eqdef \prod_{v \in M(e)}\left(\frac{1}{|N_u(v)|+1}\right)^{1/|N_u(v)|}
\end{align*}

where the empty product is considered as 1.
By Fact \ref{fact:path-prob}, $\sigma(P_{ru}) = \prod_{e : M(e) \ne \emptyset} y_e$. It is then sufficient to show 
that $y_e \le p_e$. 
Observe that $N_u(v)$ can be at most 2 since $F$ is a $3$-CNF.
For each $v \in M(e)$ with $N_u(v) =1$, the probability that $v$ does not cut $e$
is exactly $\frac{1}{2}$, independent of other nodes.
If $N_u(v)=2$ for $v\in M(e)$, $v$ marks another edge 
$e'$ along the path in addition to $e$.
The probability that $v$ cuts neither $e$
nor $e'$ is exactly $\frac{1}{2}\times\frac{2}{3}=\frac{1}{3}$, independent of other nodes.
If $N_u(v)=2$, we regard the probability of each edge surviving 
as the geometric average $\lambda$ of the survival probabilities 
of individual edges.
As a consequence, $y_e$ can be written 
as $(\frac{1}{2})^a (\frac{1}{\sqrt{3}})^b$
for some non-negative integers $a$ and $b$ 
where $a$ is the number of ancestors $v$ of $e$ such 
that $v$ marks exactly one edge along the path
and $b$ is the number of ancestors $v$ of $e$ such 
that $v$ marks exactly two edges along the path.
We now argue that $y_e$ is at most $p_e$.
If $e$ is not marked, then $a+b=0$ and hence $y_e=p_e$.
If $e$ is marked, we have $a+b>0$ which implies $y_e\leq \lambda = p_e$.
\end{proof}

The following lemma determines $M'(w,d)$.

\begin{lemma} 
\label{thm:Mwd}
For all $0\le d\le n, 0\le w\le 3d$, we have
\[
M'(w,d) = 
\begin{cases}
(2+\lambda)^{w}3^{d-w} & 0\le w \le d\\
(1+2\lambda)^{w-d}(2+\lambda)^{2d-w} & d\le w\le 2d\\
(3\lambda)^{w-2d}(1+2\lambda)^{3d-w} & 2d\le w\le 3d\\
\end{cases}
\]
\end{lemma}

\cref{thm:Mwd} already gives an upper bound $\total(T) \le M'(3t - n, t)$. 
However, taking advantage 
of \cref{pess-ntran},
we can improve this bound 
to establish
\cref{thm:main-result} for monotone formulas.  

\subsubsection{Proof of \cref{thm:main-result} for monotone formulas}

\begin{proof}
\phantomsection\label{proof:monotone-main-result}
By \cref{fact:ntrans-tree} and \cref{path-pessim} the expected time of $\ts$ is bounded by 
$\total'(T)$ (up to polynomial factors). 
We divide the proof into cases based on the value of $t$.
\begin{enumerate}[label=\textbf{Case \arabic*:}, itemindent=*, leftmargin=0em]

\item
$t\le \frac{n}{3}$. Applying \cref{pess-ntran} for the case $t\le \frac{n}{3}$, we 
get $\total(T) \le 3^t$.

\item
$\frac{n}{3} < t \le \frac{3n}{7}$.
$t \le \frac{3n}{7}$ implies $3t-n \le \frac{2t}{3}$.
We apply
\cref{pess-ntran} together with 
\cref{thm:Mwd} for the case $0\le w\le d$ to get 
\[
\total(T) \le 3^{\frac{t}{3}} M'\left(3t-n, \frac{2t}{3}\right) = \left(\frac{3}{2+\lambda}\right)^n\left(\frac{(2+\lambda)^3}{9}\right)^t \le 1.164^n \times 1.9023^t
\]

\item
$\frac{3n}{7}\le t \le \frac{n}{2}$. 
We note that $t \le \frac{n}{2}$ implies 
$3t-n \le t \le \frac{4t}{3}$ and $t \ge \frac{3n}{7}$ implies $3t-n \ge \frac{2t}{3}$. 
We apply
\cref{pess-ntran} together with 
\cref{thm:Mwd} for the case $d\le w\le 2d$ to get 
\[
\total(T) \le 3^{\frac{t}{3}} M'\left(3t-n, \frac{2t}{3}\right) = \left(\frac{2+\lambda}{1+2\lambda}\right)^n \left(\left(\frac{3(1+2\lambda)^{7}}{(2+\lambda)^{5}}\right)^{1/3}\right)^t \le 1.1962^n \times 1.7851^t
\]
\end{enumerate}
\end{proof}

\subsubsection{Proof of \cref{thm:Mwd}}

Let $T'$ be a tree of depth $d$ and weight $0\leq w\leq 3d$.
It is clear that the survival value $\total'(T')$ of $T'$ is determined once
the edges are marked consistent with the fact that every root to leaf shoot has weight at least $w$.
We say that a transversal tree $T'$ of depth $d$ and weight $w$ is \emph{normal} if it has the following marking:
Let $w=id+j$ where $0\leq i\leq 3$ and $0\leq j<d$.
Mark $i+1$ children of every non-leaf node in the 
first $j$ levels and mark $i$ children
for each of the remaining non-leaf nodes. The survival value of normal tree is exactly $((i+1)\lambda + 2 - i)^j(i\lambda + 3 - i)^{d - j}$. One can easily see that this is given exactly as $M'(w,d)$ as in the statement of the lemma.
We will show that normal trees have the largest
survival values by induction on $d$ which completes the proof
of \cref{thm:Mwd}.

Let $T'$ be a tree of depth $d$ and weight $w$ with the maximum survival value $\total'(T')$.
Let $r$ be the root of $T'$. Assume that $l_1$ children of $r$ are marked.
Let $T'_1, T'_2$, and $T'_3$ be the subtrees of $T'$ of depth $d-1$ 
and weight $w-l_1$ with $r_1, r_2$, and $r_3$ as their root nodes respectively.
$T'_1, T'_2$, and $T'_3$ are normal by induction hypothesis.
Assume that $l_2$ child edges of each of $r_i$ are marked.
The survival value of $T'$ is 
$g_1 g_2 M'(w-(l_1+l_2), d-2)$ where
$g_1= (3-l_1 +l_1 \lambda)$ and $g_2=(3-l_2 + l_2\lambda)$.
It is easy to see that if $l_1+l_2$ is held constant, $g_1 g_2$ is maximized when $l_1$ and $l_2$
are as equal as possible. If $|l_1-l_2|=1$, the survival value of $T'$ does not change if $r$ has marked $l_2$ children and 
each $r_i$ has $l_1$ marked children, that is, if the number of markings of the first two levels are exchanged.

If $l_1 = l_2$, then $T'$ is normal.
If $|l_1-l_2|\geq 2$, then $T'$ does not have the largest
survival value which is a contradiction.
We are left with the case that  $l_1$ and $l_2$ differ by one. If $l_1<l_2$, we swap $l_1$ and $l_2$ without changing the survival value and normality follows from induction. If $l_1>l_2$, the tree is already normal. Otherwise, its survival value cannot be the maximum. 
\section{Analysis of $\ts$ for arbitrary $3$-CNFs} \label{sec:non-monotone}

In this section, we analyze transversal trees for arbitrary $3$-CNFs and prove \cref{thm:main-result}. We will introduce few more ideas in addition to those introduced in \cref{sec:monotone}.

Throughout this section, we fix a $3$-CNF $F = (X, \cC)$ and let $T$ be its canonical transversal tree with root node $r$. Let the number of maximally disjoint width $3$ clauses used to develop $F$ be $m$. Let $X_D\subset X$ denote set of variables that appeared in this set of $m$ disjoint clauses. 
We also note that unlike in \cref{sec:monotone}, the clauses used to develop $T$ may not have width exactly $3$.



\subsection{Slight modification to canonical ordering and $\ts$}

We extend the conditions on the canonical ordering $\Pi$ of clauses and how $\ts$ uses $\Pi$. We order clauses in $\Pi$ so that all maximally disjoint width $3$ clauses appear first, followed by all width $3$ monotone clauses, followed by all other clauses.
For a node $u$ at level below $m$, we impose that instead of choosing the first unsatisfied monotone clause from $\Pi$, $u$ instead chooses an unsatisfied width $3$ monotone clause $C$ from $\Pi$ such that $C$ does not contain any variable $x\in X_D$ that has appeared twice in the shoot $S_{ru}$. If such a clause does not exist, then $C$ can pick the first unsatisfied monotone clause from $\Pi$.

\subsection{Fullness and Double marking}

We reuse the notion of weight here and observe that all basic facts and basic lemmas from monotone analysis apply here as well. We introduce one more definition related to this:

\begin{definition}[Uniform Weight]
Let $S = S_{uv}$ be a shoot in $T$ of length $\ell$. 
Let $a$ be the number of edges in the shoot $S_{uv}$.
The \emph{uniform weight} of $S$ denoted by $W^{+}(S) = W(S) + 3\ell - a$.
\end{definition}

We can similarly find a lower bound to this quantity for every root to leaf path:

\begin{fact}
\label{fact:shoot-uniform-extended-weight-lower-bound}
Every root to leaf shoot in $T$ has a uniform weight of
at least $3t - n$.
\end{fact}

\begin{proof}
Let $S$ be arbitrary root to leaf path with $a$ edges.
As there are only $n$ distinct edge labels, at least $a-n$ labels appear at least twice. So, $W(S) \ge a - n$.
Since the depth of $T$ is $t$, we infer that $W^{+}(S) = W^+(S) + 3t - a \ge 3t - n$ as desired.
\end{proof}

We extend the idea of markings and introduce double markings:

\begin{definition}[Double marking]
We say an edge $e\in T$ is doubly marked if $|M(e)| \ge 2$.
Let $P = P_{uv}$ be a path from $u$ to $v$. We write $W_{\ge 2}(P)$ to denote number of edges $e$ in $P$ such that $|M(e)| \ge 2$.
\end{definition}

Recall that when proving \cref{pess-ntran}, we took advantage of the fact that any monotone $3$-CNF $G$ with $\trn(G) = t$ will contain $\frac{t}{3}$ disjoint monotone clauses. However, there is no such guarantee for arbitrary $3$-CNFs. Observe that if the number of maximally disjoint monotone clauses is small, then many clauses of width $3$ will intersect with it and this will cause many edges to be doubly marked. We formalize this intution and introduce a new parameter called \emph{fullness} that will help keep track of this:

\begin{definition}[Fullness]
Let $u$ be arbitrary node at level $\ge m$ in $T$. Let $u_m$ be the node at level $m$ along the path $P_{ru}$. Then, \emph{fullness} of the shoot $S_{ru}$ is defined as
\[
\openfunction(S_{ru}) := |\{x\in X_D\setminus Q_{u_m}: \exists e = (a, b)\in S_{ru}, depth(a) \ge m, Q_e = x\}|,
\]
i.e., the number of variables in $X_D$ that are not along the path in the first $m$ levels and appear as labels of some edge in the shoot after level $m$. For arbitrary nodes $u, v$ where $u$ is ancestor of $v$ and $u$ appears at level $\ge m$, we define $\openfunction(S_{uv}) = \openfunction(S_{rv}) - \openfunction(S_{ru})$.
\end{definition}

This parameter is useful because every node after level $m$ will have at least one edge that will either `add to' fullness or at least one edge that will have marking set of size at least $2$. The edges that are `doubly marked' will contribute very little to the recursion. Moreover, we will see that fullness of any root to leaf shoot is at most $2m$. As $m$ is small, very few nodes will be such that they will not contain any doubly marked edges. This is our key insight and we fomrally prove this now.

\begin{fact}
\label{fact:disjoint-intersect}
   Let $M$ denote the set of all monotone clauses of width $3$ in $F$. Then, every clause $C\in M$ must contain at least one variable $x$ such that $x\in X_D$.
\end{fact}

\begin{lemma}
\label{lemma:disjoint-marking}
Let $u\in T$ be a node at level greater than $m$. Then, at least one of the following must be true:
\begin{enumerate}
  \item
  $u$ has at most $2$ edges going out. 

  \item
  $u$ has one edge $e$ going out such that $|M(e)| \ge 2$.

  \item
  $u$ has one edge $e$ going out such that $Q_e\in X_D$ and $|M(e)| = 1$.
\end{enumerate}
\end{lemma}

\begin{proof}
Set variables that are part of $Q_u$ to $1$ and let $F'$ be the simplified $3$-CNF.
Say case $1$ does not happen. Then, the monotone clause $C$ used to develop edges out of $u$ has width $3$ and so, $C$ is also present in $F$. By \cref{fact:disjoint-intersect}, $u$ contains an edge $e$ such that $Q_e = x$ and $x\in X_D$. If $M(e) \ge 2$, then case $2$ is satisfied and if $M(e) = 1$, then case $3$ is satisfied.
\end{proof}

\subsection{An Analysis of $\ts$ for arbitrary $3$-CNF}

We extend \cref{path-pessim} for arbitrary $3$-CNFs taking into double markings into account.

\begin{lemma}
\label{lemma:double-marking-weight}
Let $T$ be a transversal tree and let $P = P_{ru}$ be a path starting from root $r$. Then $\sigma(P) \le (\frac{1}{\sqrt{3}})^{W^{+}(P) + W_{\ge 2}(P)}$.
\end{lemma}

\begin{proof}
For a marked edge $e \in P$, we define the contribution of $e$ as
\[
q_e := \prod_{v \in M(e)}\left(\frac{1}{|N_u(v)|+1}\right)^{1/|N_u(v)|}.
\]
By Fact \ref{fact:path-prob}, $\total(P) = \prod_{e : M(e) \ne \emptyset} q_e$. It is then sufficient to show that $q_e \le \lambda$ for every marked edge $e$. Note that $q_e$ can be written as $(\frac{1}{2})^a (\frac{1}{\sqrt{3}})^b$, for some non-negative integers $a$ and $b$ such that $a + b \ge |M(e)|$. This quantity is at most $\frac{1}{\sqrt{3}}$ if $|M(e)| = 1$ and is at most $\frac{1}{3}$ if $|M(e)| \ge 2$. Finally, we observe that $e$ contributes $1$ to $W(P)$ if $|M(e)| \ge 1$ and contributes $1$ to $W_{\ge 2}(P)$ if $M(e) \ge 2$.
\end{proof}

\begin{definition}[$NM(w, d, \openinteger)$]
For non-negative integers $w, d, \openinteger$, define $NM(w, d, \openinteger)$ to be the maximum of sum of survival probabilities of leaves over depth-$d$ transversal trees $T$ for $3$-CNFs. Moreover, for every root to leaf shoot $S$, $W^+(S) \ge w$ and $\openfunction(S) \le \openinteger$.
\end{definition}





\subsubsection{Proving \cref{thm:main-result}}

We will show the following as our main lemma:

\begin{lemma}
\label{pess-nonmonotone-ntran}
Let $F$ be a $3$-CNF over $n$ variables with $\trn(F) = t \le \frac{n}{2}$. Then for any canonical transversal tree $T$ for $F$, it holds that 
\[
\total(T) \le
\begin{cases}
3^t & t \le \frac{n}{3}\\
3^{\frac{t}{3}}\times M'\left(3t - n, \frac{2t}{3}\right) & \textrm{ otherwise}
\end{cases}
\]
where $M'(w, d)$ is the same bound we obtained in \cref{sec:monotone}.
\end{lemma}

Using this, \cref{thm:main-result} follows from \cref{pess-nonmonotone-ntran} by using the exact same argument as \hyperref[proof:monotone-main-result]{in the Proof of monotone case of Theorem 1}.

Our main lemma will make use of the following bounds on $NM(w, d, \openinteger)$:

\begin{lemma} 
\label{lem:NMwdo}
For all $0\le d\le n, 0\le w\le 3d, 0\le \openinteger\le d$, it holds that:
\[
NM(w,d,\openinteger) \le
\begin{cases}
(2+\lambda)^{\openinteger}(2+\lambda^2)^{d-\openinteger}
 & 0\le w \le d\\
(2+\lambda)^{\openinteger - (w-d)}(1+2\lambda)^{w-d}(2+\lambda^2)^{d-\openinteger}
 & d\le w \le d + \openinteger\\
(1+2\lambda)^{\openinteger}(2+\lambda^2)^{2d-w}(1+\lambda+\lambda^2)^{w-d-\openinteger}
& d+\openinteger\le w\le 2d\\
(1+2\lambda)^{\openinteger - (w-2d)}(3\lambda)^{w-2d}(1+\lambda+\lambda^2)^{d-\openinteger}
& 2d\le w\le 2d+\openinteger\\
(3\lambda)^{\openinteger}(1+\lambda+\lambda^2)^{3d-w}(2\lambda + \lambda^2)^{w-2d-\openinteger}
& 2d+\openinteger\le w\le 3d\\
\end{cases}
\]
Moreover, for $\openinteger \ge d$:
\[
   NM(w, d, \openinteger) \le M'(w, d)
\]
where $M'(w, d)$ is from $\cref{sec:monotone}$.
\end{lemma}

Using this, we now prove our main lemma, which yields \cref{thm:main-result} as desired.

\begin{proof}[\textbf{Proof of \cref{pess-nonmonotone-ntran} assuming \cref{lem:NMwdo}}]
If $t \le \frac{n}{3}$, then observe that $T$ has at most $3^t$ leaves and we trivially bound $\total(T)\le 3^t$.

For $t \ge \frac{n}{3}$, we proceed by considering the maximal set of disjoint monotone width $3$ clauses in $F$ used to develop the first $m$ levels of $T$.
For every node $u\in T$ at level $m$, let the subtree rooted at $u$ be $T_u$.
We can bound $\total(T_u) \le NM(w, d, \openinteger)$ where $d = t-m, w = 3t-n, \openinteger = 2m$. Hence, $\total(T) \le 3^m NM(w, d, \openinteger)$.

If $m \ge \frac{t}{3}$, then $\openinteger = 2m \ge t - m = d$. Applying \cref{lem:NMwdo}, we infer that
\begin{align*}
\total(T) & \le 3^m M'(3t-n, t-m)\\
& = 3^{t/3} \left(3^{m-t/3} M'\left(3t-n, \frac{2t}{3} - \left(m - \frac{t}{3}\right) \right)\right)\\
& \le 3^{t/3} M'\left(3t-n, \frac{2t}{3}\right)\\
\end{align*}
and we infer the claim.

So, we assume that $m \le \frac{t}{3}$ and try to find the value of $m$ which will maximize $\total(T)$. Notice that in this case, $\openinteger \le d$ and so, we can't directly reduce to the case of $M'(w, d)$. 
As $t\le \frac{n}{2}$, it must be that $w = 3t-n \le t\le t+m \le d + \openinteger$.
This implies $w\le d+\openinteger$.
We now take two cases based on value of $w$ and apply \cref{lem:NMwdo} for the case of $\openinteger \le d$.

\begin{enumerate}[label=\textbf{Case \arabic*:}, itemindent=*, leftmargin=0em]

\item $0\le w\le d$.
In this case, we see that:
\begin{align*}
\total(T)   
   & \le 3^m NM\left(3t-n, t - m, 2m\right)\\
   & \le 3^m (2+\lambda)^{2m}(2+\lambda^2)^{t-3m}\\
   & = (2+\lambda^2)^t\left(\frac{(3)(2+\lambda)^2}{(2+\lambda^2)^3}\right)^m\\
\end{align*}
Here, the fraction has value $> 1$ and so is maximized when $m$ is maximized, i.e., when $m = \frac{t}{3}$ in which case:
\begin{align*}
\total(T)
& \le 3^{t/3} (2+\lambda)^{2t/3}\\
& = 3^{t/3} M'\left(3t-n, \frac{2t}{3}\right)
\end{align*}
Here the last equality follows by considering the case of $0\le w\le d$ for $M'(w, d)$.

\item $d\le w\le d + \openinteger$.
In this case, we see that:
\begin{align*}
\total(T)
& \le 3^m NM\left(3t-n, t - m, 2m\right)\\
& \le 3^m (2+\lambda)^{n-2t+m}(1+2\lambda)^{2t-n+m}(2+\lambda^2)^{t-3m}\\
& = \left(\frac{2+\lambda}{1+2\lambda}\right)^{n-2t}(2+\lambda^2)^t\left(\frac{(3)(2+\lambda)(1+2\lambda)}{(2+\lambda^2)^3}\right)^m\\
\end{align*}
Here, the rightmost fraction is $> 1$ and so is maximized when $m$ is maximized, i.e., when $m = \frac{t}{3}$ in which case:
\begin{align*}
\total(T)
& \le 3^{t/3} (2+\lambda)^{n - 5t/3} (1+2\lambda)^{7t/3 - n}\\
& = 3^{t/3} M'\left(3t-n, \frac{2t}{3}\right)
\end{align*}
Here the last equality follows by considering the case of $d\le w\le 2d$ for $M'(w, d)$ (we can do this as $\openinteger \le d$ and hence, $w\le 2d$).
\end{enumerate}

Thus, in either case, we exactly recover the monotone bound as desired.
\end{proof}

\subsubsection{Upper bounds on $NM(w, d, \openinteger)$}

As done in monotone analysis, we let $\lambda \eqdef \frac{1}{\sqrt{3}}$. 
Our goal is to prove \cref{lem:NMwdo}.
We introduce a recurrence relation $L(w, d, \openinteger)$ that we argue will upper bound $NM(w, d, \openinteger)$.

\begin{definition}[$L(w, d, \openinteger)$]
\label{def:Lwdo}
We define $L(w, d, \openinteger): \N^3\rightarrow \R$ recursively as follows:
\[
L(w, 0, \openinteger) = 
\begin{cases}
1 & w \le 0\\
0 & w > 0\\
\end{cases}
\]
For $w\le 3d, 0\le d\le n$, and $\openinteger \ge 1$, define $L(w, d, \openinteger)$ as:
\begin{align*}
L(w, d, \openinteger) = 
\max\{
& (2+\lambda) L(w-1, d-1, \openinteger-1),\\
& (1+2\lambda) L(w-2, d-1, \openinteger-1),\\
& 3\lambda L(w-3, d-1, \openinteger-1)
\}\\
\end{align*}
For $w\le 3d, 0\le d\le n$, and $\openinteger = 0$, define $L(w, d, 0)$ as:
\begin{align*}
L(w, d, 0) =
\max\{
& (2+\lambda^2) L(w-1, d-1, 0),\\
& (1+\lambda+\lambda^2) L(w-2, d-1, 0),\\
& (2\lambda + \lambda^2) L(w-3, d-1, 0)
\}\\
\end{align*}
\end{definition}

We claim that $L(w, d, \openinteger)$ gives a good bound on $M(w, d, \openinteger)$.

\begin{prop} 
\label{claim:NMwd-recurrence}
For all $0\le d\le n, 0\le w\le 3d, 0\le \openinteger$, it holds that:
$NM(w, d, \openinteger) \le L(w, d, \openinteger)$
\end{prop}

We will show the following bound on $L(w, d, \openinteger)$.

\begin{lemma} 
\label{lem:Lwdo}
For all $0\le d\le n, 0\le w\le 3d, 0\le \openinteger\le d$, it holds that:
\[
L(w,d,\openinteger) \le
\begin{cases}
(2+\lambda)^{\openinteger}(2+\lambda^2)^{d-\openinteger}
 & 0\le w \le d\\
(2+\lambda)^{\openinteger - (w-d)}(1+2\lambda)^{w-d}(2+\lambda^2)^{d-\openinteger}
 & d\le w \le d + \openinteger\\
(1+2\lambda)^{\openinteger}(2+\lambda^2)^{2d-w}(1+\lambda+\lambda^2)^{w-d-\openinteger}
& d+\openinteger\le w\le 2d\\
(1+2\lambda)^{\openinteger - (w-2d)}(3\lambda)^{w-2d}(1+\lambda+\lambda^2)^{d-\openinteger}
& 2d\le w\le 2d+\openinteger\\
(3\lambda)^{\openinteger}(1+\lambda+\lambda^2)^{3d-w}(2\lambda + \lambda^2)^{w-2d-\openinteger}
& 2d+\openinteger\le w\le 3d\\
\end{cases}
\]
Moreover, for $\openinteger \ge d$:
\[
   L(w, d, \openinteger) \le M'(w, d)
\]
where $M'(w, d)$ is from $\cref{sec:monotone}$.
\end{lemma}

Combining \cref{claim:NMwd-recurrence} and \cref{lem:Lwdo}, \cref{lem:NMwdo} trivially follows.

\subsubsection{Proving $NM(w, d, \openinteger)\le L(w, d, \openinteger)$}

Using \cref{lemma:disjoint-marking} and \cref{lemma:double-marking-weight}, we come up with a recurrence for $NM(w, d, \openinteger)$ and show it's bounded by $L(w, d, \openinteger)$, proving \cref{claim:NMwd-recurrence}.

\begin{proof}[\textbf{Proof of \cref{claim:NMwd-recurrence}}]
Recall that in the canonical ordering, all width $3$ monotone clauses appear first and remaining clauses appear later. After exhausting the width $3$ monotone clauses, the remaining clauses that we develop in the transversal tree have width at most $2$. Towards this, for non-negative integers $w, d$: let $M_2(w, d)$ be the maximum sum of survival probabilities over all transversal trees for $2$-CNFs where every root to leaf path has uniform weight at least $w$. Recall that uniform weight is defined with respect to $3$-CNFs and we continue using that definition.
We get various recurrences by considering cases on number of marked edges out of the root node ($0$ or $1$ or $2$) and by observing that some cases are dominated by others (such as various cases of width $1$ clauses).
The remaining recurrences that are not dominated by any other recurrence are the following:
\begin{align*}
M_2(w, d) \le 
\max\{
& (2)M_2(w-1, d-1),\\
& (1+\lambda)M_2(w-2, d-1),\\
& 2\lambda M_2(w-3, d-1)
\}\\
\end{align*}
By induction, we infer that $M_2(w, d) \le L(w, d, 0)$.

We now develop a recurrence for $NM(w, d, 0)$.
Recall that in canonical ordering, either we exhaust all width $3$ monotone clause and reach $M_2(w, d)$, or we develop width $3$ monotone clause.
Observe that \cref{lemma:disjoint-marking} guarantees that every node in such a tree must have an edge $e$ coming out of it such that $|M(e)| \ge 2$. Taking cases on the number of marked edges coming out of the root node and whether root node has $3$ or at most $2$ edges coming out, we get many recurrences.
However certain recurrences are dominated by others and the remaining recurrences that are not dominated by any othe recurrence are as follows:
\begin{align*}
NM(w, d, 0) \le 
\max\{
& (2 + \lambda^2)NM(w-1, d-1, 0),\\
& (1+\lambda+\lambda^2)NM(w-2, d-1, 0),\\
& (2\lambda+\lambda^2) NM(w-3, d-1, 0),\\
& M_2(w, d)
\}\\
\end{align*}
By induction, we again infer that $NM(w, d, 0) \le L(w, d, 0)$.

We now develop a recurrence for $NM(w, d, \openinteger)$.
We again take advantange of the fact that in canonical ordering, either we exhaust all width $3$ monotone clause and reach $M_2(w, d)$, or we develop width $3$ monotone clause. Moreover, amongst width $3$ clauses, canonical ordering causes either $\openinteger$ to decrease by at least $1$ or we exhaust such clauses and all remaining clauses have the property that a node developed using such a clause will have an outgoing edge $e$ such that $|M(e)| \ge 2$. 

We get many recurrences for $NM(w, d, \openinteger)$ by considering cases on number of marked edges ($1$ or $2$ or $3$) out of the root node, number of marked edges that cause $\openfunction$ to decrease ($1$ or $2$ or $3$), various combinations of number of double marked edges ($1$ or $2$ or $3$), whether the root node has at most $2$ edges coming out, and whether the root node has no edges that cause $\openfunction$ to decrease by at least $1$. Notice that if the root node has no edges that cause $\openfunction$ to decrease by at least $1$, then by clause ordering, no other width $3$ clause can cause $\openfunction$ to decrease and hence, we are in case $NM(w, d, 0)$. Lastly, we observe that certain recurrences are dominated by others. The remaining recurrences that are not dominated by any other recurrences are the following:
\begin{align*}
NM(w, d, \openinteger) \le 
\max\{
& (2+\lambda) NM(w-1, d-1, \openinteger-1),\\
& (1+2\lambda) NM(w-2, d-1, \openinteger-1),\\
& 3\lambda NM(w-3, d-1, \openinteger-1),\\
& NM(w, d, 0),\\
& M_2(w, d)
\}\\
\end{align*}
By induction, utilizing the fact that $L(w, d, \openinteger) \ge L(w, d, 0)$, we again infer that $NM(w, d, \openinteger) \le L(w, d, \openinteger)$ as desired.
\end{proof}

\subsubsection{Upper bound on $L(w, d, 0)$}

We first show \cref{lem:Lwdo} for the special case of $\openinteger = 0$:

\begin{lemma} 
\label{lem:Lwdo-0}
For all $0\le d\le n, 0\le w\le 3d$, it holds that:
\[
L(w,d,0) \le
\begin{cases}
(2+\lambda^2)^d & 0\le w \le d\\
(2+\lambda^2)^{2d-w}(1+\lambda+\lambda^2)^{w-d} & d\le w\le 2d\\
(1+\lambda+\lambda^2)^{3d-w}(2\lambda+\lambda^2)^{w-2d} & 2d\le w\le 3d\\
\end{cases}
\]
\end{lemma}

\begin{proof}

Let $G_1, G_2, G_3, G: \N^2\rightarrow \R$ be defined as:

\begin{eqnarray*}
G_1(w, d) & = & (2+\lambda^2)^d\\
G_2(w, d) & = & (2+\lambda^2)^{2d-w}(1+\lambda+\lambda^2)^{w-d}\\
G_3(w, d) & = & (1+\lambda+\lambda^2)^{3d-w}(2\lambda+\lambda^2)^{w-2d}\\
G(w, d) & = & \min\{G_1(w, d), G_2(w, d), G_3(w, d)\}\\
\end{eqnarray*}

For $1\le i\le 3$, define $P_i$ to be the set of pairs $(w, d)$ such that $d \ge 0$ and $w\in [(i-1)d, id]$. We will show the following two propositions:

\begin{proposition}
\label{prop:G}
For all $1\le i\le 3$, and all $(w, d)\in P_i: G(w, d) = G_i(w, d)$.
\end{proposition}

\begin{proposition}
\label{prop:L-sol-open-0}
For all $1\le i\le 3$ and all $(w, d)\in P_i: L(w, d, 0) \le G(w, d)$.
\end{proposition}

We observe that \cref{prop:G} and \cref{prop:L-sol-open-0} together imply our claim.

\begin{proof}[\textbf{Proof of \cref{prop:G}}]
The result follows immediately from the following claims:
\begin{claim}
\label{claim:G-sol-1}
$G_1(w,d) \leq G_2(w,d)$ if and only if $w \leq d$, with equality when $w=d$.
\end{claim}

\begin{claim}
\label{claim:G-sol-2}
$G_2(w,d) \leq G_3(w,d)$ if and only if $w \leq 2d$, with equality when $w=2d$.
\end{claim}

\cref{claim:G-sol-1} holds because:

\begin{equation*}
\frac{G_1(w, d)}{G_2(w, d)} = \left(\frac{2+\lambda^2}{1+\lambda+\lambda^2}\right)^{w-d}
\end{equation*}

which is greater than 1 if and only if $w>d$.

\cref{claim:G-sol-2} holds because:

\begin{equation*}
\frac{G_2(w, d)}{G_3(w, d)} = \left(\frac{(1+\lambda+\lambda^2)^2}{(2\lambda + \lambda^2)(2+\lambda^2)}\right)^{w-2d}
\end{equation*}
which is greater than 1 if and only if $w>2d$.
\end{proof}

\begin{proof}[\textbf{Proof of \cref{prop:L-sol-open-0}}]
We consider cases on value of $w$ and in every case, induct on $d$ and apply \cref{def:Lwdo} to infer the claim.

\begin{enumerate}[label=\textbf{Case \arabic*:}, itemindent=*, leftmargin=0em]

\item Assume $(w, d)\in P_1$.
\begin{eqnarray*}
L(w,d,0) & \le & \max \{  (2+\lambda^2) G(w-1,d-1,0),\\
          &   &          (1+\lambda+\lambda^2) G(w-2,d-1,0),\\
          &   &          (2\lambda+\lambda^2) G(w-3,d-1,0) \}\\
& \le  & \max\{(2+\lambda^2)G_1(w-1,d-1),(1+\lambda+\lambda^2)G_1(w-2,d-1),(2\lambda+\lambda^2)G_1(w-3,d-1)\}\\
& = & G_1(w,d) \max\{1, (1+\lambda+\lambda^2)/(2+\lambda^2), (2\lambda+\lambda^2)/(2+\lambda^2)\}\\
& = & G_1(w,d)\\
& = & G(w, d)
\end{eqnarray*}
The last equality follows by applying \cref{prop:G} for the case $(w, d)\in P_1$.

\item Assume $(w, d)\in P_2$.
\begin{eqnarray*}
L(w,d,0) & \le & \max\{(2+\lambda^2) G(w-1,d-1,0),\\
          &   &        (1+\lambda+\lambda^2) G(w-2,d-1,0),\\
          &   &       (2\lambda+\lambda^2) G(w-3,d-1,0) \} \\
& \le & \max \{(2+\lambda^2)G_2(w-1,d-1),(1+\lambda+\lambda^2)G_2(w-2,d-1),(2\lambda+\lambda^2)G_2(w-3,d-1)\}\\
& = & G_2(w,d) \max\{1, 1, (2\lambda+\lambda^2)(2+\lambda^2)/(1+\lambda+\lambda^2)^2\}\\
& = & G_2(w,d)\\
& = & G(w, d)
\end{eqnarray*}
The last equality follows by applying \cref{prop:G} for the case $(w, d)\in P_2$.

\item Assume $(w, d)\in P_3$.
\begin{eqnarray*}
L(w,d,0) & \le & \max\{(2+\lambda^2) G(w-1,d-1,0),\\
         &   &        (1+\lambda+\lambda^2) G(w-2,d-1,0),\\
         &   &        (2\lambda+\lambda^2) G(w-3,d-1,0) \}\\
& \le & \max \{(2+\lambda^2)G_3(w-1,d-1),(1+\lambda+\lambda^2)G_3(w-2,d-1),(2\lambda+\lambda^2)G_3(w-3,d-1)\}\\
& = & G_3(w,d) \max\{(2+\lambda^2)(2\lambda+\lambda^2)/(1+\lambda+\lambda^2)^2, 1, 1\}\\
& = & G_3(w,d)\\
& = & G(w,d)\\
\end{eqnarray*}
The last equality follows by applying \cref{prop:G} for the case $(w, d)\in P_3$.

\end{enumerate}

\end{proof}

\end{proof}

\subsubsection{Upper bound on $L(w, d, \openinteger)$}

We are finally ready to give general bounds on $L(w, d, \openinteger)$:

\begin{proof}[\textbf{Proof of \cref{lem:Lwdo}}]

Notice that if $\openinteger \ge d$, then if we try and unravel the recurrence, no path can lead to the case $\openinteger = 0, d > 0$. Hence, $\openinteger$ plays no role in restricting the recurrence and $L(w, d, \openinteger)$ follows the same recurrence as $M'(w, d)$, yielding the claim.

For $\openinteger \le d$, we proceed by first defining $H_1, H_2, H_3, H_4, H_5, H: \N^3\rightarrow \R$ as follows:

\begin{eqnarray*}
H_1(w, d, \openinteger) & = & (2+\lambda)^{\openinteger}(2+\lambda^2)^{d-\openinteger}\\
H_2(w, d, \openinteger) & = & (2+\lambda)^{\openinteger - (w-d)}(1+2\lambda)^{w-d}(2+\lambda^2)^{d-\openinteger}\\
H_3(w, d, \openinteger) & = & (1+2\lambda)^{\openinteger}(2+\lambda^2)^{2d-w}(1+\lambda+\lambda^2)^{w-d-\openinteger}\\
H_4(w, d, \openinteger) & = & (1+2\lambda)^{\openinteger - (w-2d)}(3\lambda)^{w-2d}(1+\lambda+\lambda^2)^{d-\openinteger}\\
H_5(w, d, \openinteger) & = & (3\lambda)^{\openinteger}(1+\lambda+\lambda^2)^{3d-w}(2\lambda + \lambda^2)^{w-2d-\openinteger}\\
H(w, d, \openinteger) & = & \min\{H_2(w, d, \openinteger), H_3(w, d, \openinteger), H_4(w, d, \openinteger), H_5(w, d, \openinteger)\}\\
\end{eqnarray*}

For $1\le i\le 5$, define $Q_i\subset \N^3$ as follows:

\begin{align*}
Q_1 & = \{(w, d, \openinteger)\in \N^3: 0 \le w \le d + \openinteger\}\\
Q_2 & = \{(w, d, \openinteger)\in \N^3: d \le w \le d + \openinteger\}\\
Q_3 & = \{(w, d, \openinteger)\in \N^3: d + \openinteger \le w \le 2d\}\\
Q_4 & = \{(w, d, \openinteger)\in \N^3: 2d \le w \le 2d + \openinteger\}\\
Q_5 & = \{(w, d, \openinteger)\in \N^3: 2d + \openinteger \le w \le 3d\}\\
\end{align*}

 We will show the following propositions that together imply our claim:

\begin{proposition}
\label{prop:H}
For all $1\le i\le 5$, and all $(w, d, \openinteger)\in Q_i: H(w, d, \openinteger) = H_i(w, d, \openinteger)$.
\end{proposition}

\begin{proposition}
\label{prop:L-sol-open}
For all $1\le i\le 5$ and all $(w, d, \openinteger)\in Q_i: L(w, d, \openinteger) \le H(w, d, \openinteger)$.
\end{proposition}

We will in fact use \cref{prop:H} in the proof of \cref{prop:L-sol-open}. Hence, we prove the former first:

\begin{proof}[\textbf{Proof of \cref{prop:H}}]
The result follows immediately from the following claims:

\begin{claim}
\label{claim:H-sol-1}
$H_1(w, d, \openinteger) \leq H_2(w, d, \openinteger)$ if and only if $w \leq d$, with equality when $w = d$.
\end{claim}

\begin{claim}
\label{claim:H-sol-2}
$H_2(w, d, \openinteger) \leq H_3(w, d, \openinteger)$ if and only if $w \leq d + \openinteger$, with equality when $w = d + \openinteger$.
\end{claim}

\begin{claim}
\label{claim:H-sol-3}
$H_3(w, d, \openinteger) \leq H_4(w, d, \openinteger)$ if and only if $w \leq 2d$, with equality when $w = 2d$.
\end{claim}

\begin{claim}
\label{claim:H-sol-4}
$H_4(w, d, \openinteger) \leq H_5(w, d, \openinteger)$ if and only if $w \leq 2d+\openinteger$, with equality when $w = 2d+\openinteger$.
\end{claim}

\cref{claim:H-sol-1} holds because: 

\begin{equation*}
\frac{H_1(w, d, \openinteger)}{H_2(w, d, \openinteger)} = \left(\frac{2+\lambda}{1+2\lambda}\right)^{w-d}
\end{equation*}

which is greater than 1 if and only if $w > d$.

\cref{claim:H-sol-2} holds because: 

\begin{equation*}
\frac{H_2(w, d, \openinteger)}{H_3(w, d, \openinteger)} = \left(\frac{(1+2\lambda)(2+\lambda^2)}{(2+\lambda)(1+\lambda+\lambda^2)}\right)^{w-d-\openinteger}
\end{equation*}

which is greater than 1 if and only if $w > d + \openinteger$.

\cref{claim:H-sol-3} holds because:

\begin{equation*}
\frac{H_3(w, d, \openinteger)}{H_4(w, d, \openinteger)} = \left(\frac{(1+2\lambda)(1+\lambda+\lambda^2)}{(2+\lambda^2)(3\lambda)}\right)^{w-2d}
\end{equation*}

which is greater than 1 if and only if $w>2d$.

\cref{claim:H-sol-4} holds because:

\begin{equation*}
\frac{H_4(w, d, \openinteger)}{H_5(w, d, \openinteger)} =\left(\frac{(3\lambda)(1+\lambda+\lambda^2)}{(1+2\lambda)(2\lambda+\lambda^2)}\right)^{w-2d-\openinteger}
\end{equation*}

which is greater than 1 if and only if $w>2d+\openinteger$.
\end{proof}

We prove our final proposition:

\begin{proof}[\textbf{Proof of \cref{prop:L-sol-open}}]
We observe that for $\openinteger = 0$, our claim follows from \cref{lem:Lwdo-0}.
We use this fact in the inductive argument below and only consider cases where $\openinteger \ge 1$. We consider cases on value of $w$ and in every case, induct on $d + \openinteger$ and apply \cref{def:Lwdo} to infer the claim.  

\begin{enumerate}[label=\textbf{Case \arabic*:}, itemindent=*, leftmargin=0em]

\item Assume $(w, d, \openinteger)\in Q_1$ and $\openinteger \ge 1$.
\begin{eqnarray*}
L(w, d, \openinteger) & = & \max\{(2+\lambda)H(w-1, d-1, \openinteger-1), (1+2\lambda)H(w-2, d-1, \openinteger-1), \\
& & (3\lambda)H(w-3, d-1, \openinteger-1)\}\\
& \le & \max\{(2+\lambda)H_1(w-1, d-1, \openinteger-1), (1+2\lambda)H_1(w-2, d-1, \openinteger-1), \\
& & (3\lambda)H_1(w-3, d-1, \openinteger-1)\}\\
& = & H_1(w, d, \openinteger)\max\left\{1, \frac{1+2\lambda}{2+\lambda}, \frac{3\lambda}{2+\lambda}\right\} \\
& = & H_1(w, d, \openinteger)\\
& = & H(w, d, \openinteger)
\end{eqnarray*}
The last equality follows by applying \cref{prop:H} for the case $(w, d,\openinteger)\in Q_1$.

\item Assume $(w, d, \openinteger)\in Q_2$ and $\openinteger \ge 1$.
\begin{eqnarray*}
L(w, d, \openinteger) & \leq & \max\{(2+\lambda)L(w-1, d-1, \openinteger-1), (1+2\lambda)L(w-2, d-1, \openinteger-1), \\
& & (3\lambda)L(w-3, d-1, \openinteger-1)\}\\
& \le & \max\{(2+\lambda)H_2(w-1, d-1, \openinteger-1), (1+2\lambda)H_2(w-2, d-1, \openinteger-1), \\
& & (3\lambda)H_2(w-3, d-1, \openinteger-1)\}\\
& = & H_2(w, d, \openinteger)\max\left\{1, 1, \frac{(3\lambda)(2+\lambda)}{(1+2\lambda)^2} \right\}\\
& = & H_2(w, d, \openinteger)\\
& = & H(w, d, \openinteger)\\
\end{eqnarray*}
The last equality follows by applying \cref{prop:H} for the case $(w, d,\openinteger)\in Q_2$.

\item Assume $(w, d, \openinteger)\in Q_3$ and $\openinteger \ge 1$.
\begin{eqnarray*}
L(w, d, \openinteger) & \leq & \max\{(2+\lambda)L(w-1, d-1, \openinteger-1), (1+2\lambda)L(w-2, d-1, \openinteger-1), \\
& & (3\lambda)L(w-3, d-1, \openinteger-1)\}\\
& \le & \max\{(2+\lambda)H_3(w-1, d-1, \openinteger-1), (1+2\lambda)H_3(w-2, d-1, \openinteger-1), \\
& & (3\lambda)H_3(w-3, d-1, \openinteger-1)\}\\
& = & H_3(w, d, \openinteger)\max\left\{\frac{(2+\lambda)(1+\lambda+\lambda^2)}{(1+2\lambda)(2+\lambda^2)}, 1, \frac{(2+\lambda^2)(3\lambda)}{(1+2\lambda)(1+\lambda+\lambda^2)} \right\}\\
& = & H_3(w, d, \openinteger)\\
& = & H(w, d, \openinteger)\\
\end{eqnarray*}
The last equality follows by applying \cref{prop:H} for the case $(w, d,\openinteger)\in Q_3$.

\item Assume $(w, d, \openinteger)\in Q_4$ and $\openinteger \ge 1$.
\begin{eqnarray*}
L(w, d, \openinteger) & \leq & \max\{(2+\lambda)L(w-1, d-1, \openinteger-1), (1+2\lambda)L(w-2, d-1, \openinteger-1), \\
& & (3\lambda)L(w-3, d-1, \openinteger-1)\}\\
& \le & \max\{(2+\lambda)H_4(w-1, d-1, \openinteger-1), (1+2\lambda)H_4(w-2, d-1, \openinteger-1), \\
& & (3\lambda)H_4(w-3, d-1, \openinteger-1)\}\\
& = & H_4(w, d, \openinteger)\max\left\{\frac{(3\lambda)(2+\lambda)}{(1+2\lambda)^2}, 1, 1 \right\}\\
& = & H_4(w, d, \openinteger)\\
& = & H(w, d, \openinteger)
\end{eqnarray*}
The last equality follows by applying \cref{prop:H} for the case $(w, d,\openinteger)\in Q_4$.

\item Assume $(w, d, \openinteger)\in Q_5$ and $\openinteger \ge 1$.
\begin{eqnarray*}
L(w, d, \openinteger) & \leq & \max\{(2+\lambda)L(w-1, d-1, \openinteger-1), (1+2\lambda)L(w-2, d-1, \openinteger-1), \\
& & (3\lambda)L(w-3, d-1, \openinteger-1)\}\\
& \le & \max\{(2+\lambda)H_5(w-1, d-1, \openinteger-1), (1+2\lambda)H_5(w-2, d-1, \openinteger-1), \\
& & (3\lambda)H_5(w-3, d-1, \openinteger-1)\}\\
& = & H_5(w, d, \openinteger)\max\left\{\frac{(2+\lambda)(2\lambda+\lambda^2)^2}{(3\lambda)(1+\lambda+\lambda^2)^2},
                             \frac{(1+2\lambda)(2\lambda+\lambda^2)}{(3\lambda)(1+\lambda+\lambda^2)}, 1 \right\}\\
& = & H_5(w, d, \openinteger)\\
& = & H(w, d, \openinteger)
\end{eqnarray*}
The last equality follows by applying \cref{prop:H} for the case $(w, d,\openinteger)\in Q_5$.

\end{enumerate}

\end{proof}

\end{proof}

\section{Satisfiability for CNFs with bounded negations} \label{sec:satboundedneg}

We now use $\ts$ to give an enumeration algorithm for class of CNFs with arbitrary width and bounded negations in each clause.

We will use the following well known estimate of binomial coefficients:
\begin{proposition}
Let $H_2: (0, 1) \rightarrow (0, 1)$ be the binary entropy function defined as $H_2(x) = -x \log_2(x) + (1-x) \log_2(1-x)$. Then, for $k\le n / 2$, it holds that: $\sum_{i = 0}^k\binom{n}{k} \le \poly(n) 2^{n H_2(k / n)}$.
\end{proposition}

\begin{proof}[\textbf{Proof of \cref{thm:enum3negate}}]
Without loss of generality we assume each clause $F$ contains at most $3$ postive literals. Indeed, if every clause in $F$ contains at most $3$ negative literals, then we can negate every literal in every clause and consider the resultant CNF. This CNF is satisfiable if and only if the original CNF was satisfiable. Moreover, the new CNF has the property that each clause contains at most $3$ positive literals.

Let $c = 0.71347$.
Then, we use $\ts$ to to enumerate all minimal satisfiable assignments of weight at most $cn$. We then exhaustively go over all assignments $\alpha$ with weight at least $cn$ and check whether $\alpha$ satisfies $F$ and output such minimal $\alpha$.

The runtime of the exhaustive procedure is
\[
	\poly(n)\sum_{i=cn}^n \binom{n}{k} \le \poly(n) 2^{n H_2(c)} \le O(1.8204^n)
\]

Notice that when we develop the transversal tree, we only develop positive monotone clauses. Any positive monotone clauses that we encounter during the $\ts$ procedure for $F$ must have width at most $3$ as each clause contains at most $3$ positive literals. Hence, the resultant transversal tree $T$ is still a ternary tree. So, every root to leaf shoot $S$ must have weight at least $3t - n$ where $t = cn$. We do not put any lower bound on $\openfunction$ for any such shoot and so, we set $\openinteger = \infty$. Then, the runtime of $\ts$ upto polynomial factors is bounded by $NM(3(cn) - n, cn, \infty) \le M'((3c-1), cn)$. We observe that $c\le 3c-1 \le 2c$ and so, we are in the regime where $w\le d\le 2d$. Thus, 
\begin{align*}
M'((3c-1)n, cn)
& \le \left(\left( 1 + \frac{2}{\sqrt{3}}\right)^{2c-1}\left(2 + \frac{1}{\sqrt{3}}\right)^{1-c}\right)^n\\
& \le 1.8204^n
\end{align*}
Hence, the runtime of our algorithm is indeed $O(1.8204^n)$ as desired.

\end{proof}

\section{Conclusion} \label{sec:conclusion}

We gave a new non-trivial algorithm for \enumballsat{3}{\frac{n}{2}}: given an $n$-variable 3-CNF with no satsifying assignment of Hamming weight less than $\frac{n}{2}$, we can enumerate all satisfying assignments of Hamming weight exactly $\frac{n}{2}$ in expected time $1.598^n$. Several fascinating questions with major consequences remain open. Here we list the most pressing.

\begin{enumerate}
	\item We already mentioned that \enumballsat{3}{\frac{n}{2}} cannot be solved in less than $1.565^n$ steps. Close this gap.


	\item Can our approach produce significant improvements for $k$-CNFs with $k>3$?

\end{enumerate}

It seems that to make progress towards resolving these problems, deeper analysis of the structure of $k$-CNFs will be required. 

\bibliographystyle{alpha}
\bibliography{refs}

\end{document}